# A study on the universality of the magnetic-field-induced phase transitions in the two-dimensional electron system in an AlGaAs/GaAs heterostructure

C. F. Huang[1,2], Y. H. Chang[1,*], H. H. Cheng[3], C. -T. Liang[1], and G. J. Hwang[2]

[1]Department of Physics, National Taiwan University, Taipei 106, Taiwan, R. O. C.

[2]National Measurement Laboratory, Center for Measurement Standards, Industrial Technology Research Institute, Hsinchu 300, Taiwan, R. O. C.

[3]Center for Condensed Matter Sciences, National Taiwan University, Taipei 106, Taiwan, R. O. C.

**abstract**

Plateau-plateau (P-P) and insulator-quantum Hall conductor (I-QH) transitions are observed in the two-dimensional electron system in an AlGaAs/GaAs heterostructure. At high fields, the critical conductivities are not of the expected universal values and the temperature-dependence of the width of the P-P transition does not follow the universal scaling. However, the semicircle law still holds, and universal scaling behavior was found in the P-P transition after mapping it to the I-QH transition by the Landau-level addition transformation. We pointed out that in order to get a correct critical exponent, it is essential that the scaling analysis must be performed near the critical point. And with proper analysis, we found that the P-P transition and the insulator quantum Hall conductor transitions are of the same universal class.

[*] Corresponding author. Tel.: 0-886-2-3366-5126; fax: 0-886-2-3366-5126; e-mail: yhchang@phys.ntu.edu.tw.

Two kinds of magnetic-field-induced phase transitions could be observed in the quantum Hall effect when a magnetic field *B* is applied perpendicular to a two-dimensional (2D) system. The first type is the plateau-plateau (P-P) transitions that occur between two adjacent quantum Hall plateaus, [1-5] and the second type that could be observed is the insulator-quantum Hall conductor (I-QH) transitions. [1-3,6-8] These two transitions are expected to be of the same universal class governed by the same semicircle law and universal scaling. [1-4,9,10]. The universality of critical conductivities, i.e. the longitudinal and Hall conductivities $\sigma_{xx}$ and $\sigma_{xy}$ at the critical point, are also expected. [2,11] In fact, all magnetic-field-induced phase transitions are equivalent according to the law of corresponding states, by which S. Kivelson et al. [1] construct the global phase diagram (GPD) of the quantum Hall effect. In the law of corresponding states the equivalence between different transitions is established by the Landau-level addition transformation. [1-3] However, the universality of magnetic-field-induced phase transitions has been a subject of debate for many years. [3-7,12-19] Recently, we [3] showed that the equivalence between different magnetic-field-induced phase transitions can be found by suitable analysis even when some expected properties are broken. [3]

To further study the universality of magnetic-field-induced phase transitions, we performed a magneto-transport study on the two-dimensional electron system (2DES) in an AlGaAs/GaAs heterostructure. The sample used in this study is an AlGaAs/GaAs heterostructure grown by MBE. The heterostructure consists of a 0.5 $\mu$m thick GaAs layer grown on a semi-insulating GaAs substrate, a 100 Å thick undoped $Al_{0.22}Ga_{0.78}As$ spacer, a 300 Å thick Si-doped $Al_{0.22}Ga_{0.78}As$ layer with doping concentration $8\times 10^{17}$/cm$^3$, a 100 Å thick undoped Al$_{0.22}$Ga$_{0.78}$As layer, and a 100 Å thick Si-doped GaAs cap layer with doping concentration $8\times 10^{17}$/cm$^3$. The sample is made into the Hall pattern by standard lithography and etching processes, and indium was alloyed into contact regions at 450ºC in N$_2$ atmosphere to form Ohmic contacts. Al was deposited on the surface of the Hall bar to serve as the Schottky gate. With a top-loading He$^3$ system and a 15 T superconductor magnet, magneto-transport measurements were performed by the low-frequency AC lock-in technique. In this study, the gate voltage $V_g$=-0.1V, and the AC current applied to the sample is 0.1 $\mu$A. From the oscillations of the longitudinal resistivity $\rho_{xx}$ and the Hall measurements at low $B$, the carrier concentration $n$=2.0×10$^{11}$/cm$^2$. The mobility $\mu$ at $T$=0.31K obtained from $\rho_{xx}|_{B=0}$=1/$ne\mu$ is 7.6×10$^3$ cm$^2$/Vs.

Figure 1 shows the curves of $\rho_{xx}$ and the Hall resistivity $\rho_{xy}$. As reported in Ref. 20, this sample is an insulator at low fields and makes an I-QH transition inconsistent with the GPD to enter the quantum Hall liquid state. At high fields, in Fig. 1 there are well-developed $\nu$=2 and 1 quantum Hall states in which $\rho_{xx}$ approaches zero and $\rho_{xy}$ remains $h/\nu e^2$, and the 2DES is an insulator when $B$ is larger than 14.7T. In this paper, we focus on the P-P transitions between the $\nu$=2 and 1 quantum Hall states and the I-QH transition between the $\nu$=1 quantum Hall state and the high-field insulating state, where ν is the filling factor. For convenience, we denote the observed P-P and I-QH transitions as the 2-1 and 1-0 transitions with the numbers 2, 1, and 0 present the quantum Hall states of ν=2 and 1 and the insulating states, respectively.

In Fig. 1, $\rho_{xy}$ remains $h/e^2$ at the lowest temperature $T$=0.31 K not only in the $\nu$=1 quantum Hall state, but also in the insulating region where $B > B_c$=14.7T and hence the semicircle law holds. [2,10] Here $B_c$ is the critical magnetic field. In an I-QH transition, usually the scaling is studied by examining the collapse of $\rho_{xx}$ with respect to the scaling parameter $(\nu-\nu_c)/T^\kappa$ or by examining the $T$-dependence of the slope $d\rho_{xx}/dB$ at $B_c$. [2,13,16,20,21] Here $\nu_c$ is filling factor at the critical point. As shown in the inset of Fig. 1, in the 1-0 transition the curves of $\rho_{xx}$ at different temperatures collapse with respect to $(\nu-\nu_c)/T^\kappa$ when we take $\kappa=0.38$, which is close to the expected universal value $0.42\pm0.04$ [2-5]. Therefore, in our study the 1-0 transition satisfies both the semicircle law and universal scaling. However, at the critical point $\rho_{xx}\sim37$k$\Omega$ rather than the expected value $h/e^2\sim25.8$ k$\Omega$, which indicates the failure of the universality of critical conductivities. [2,3] In a P-P transition, conventionally the critical exponent $\kappa$ is obtained from [3-5,17]

$$\ln(\max|d\rho_{xy}/dB|) = \kappa \ln T + \text{const} \qquad (1)$$

because the transition width in $\rho_{xy}$ can be approximated by the inverse of $\max|d\rho_{xy}/dB|$ and should be proportional to $T^\kappa$ from scaling. However, in the 2-1 P-P transition, using such analysis, we obtained $\kappa=0.64\pm0.09$ from the slope of $\ln(\max|d\rho_{xy}/dB|)$-$\ln T$ and hence the universal scaling is invalid.

We showed earlier that in the IQHE, a P-P transition can still be mapped to the 1-0 I-QH transition by the Landau-level addition transformation even when the universal scaling is not observed. [3] In the 2-1 transition, below the Fermi energy there is one filled Landau band, which contributes $e^2/h$ to $\sigma_{xy}$ and has no contribution to $\sigma_{xx}$. [1-3] To map the 2-1 transition to the 1-0 I-QH transition, we define

$$\sigma^t_{xy} = \sigma_{xy} - e^2/h \quad \text{and} \quad \sigma^t_{xx} = \sigma_{xx} \qquad (2)$$

to subtract the contribution of this filled Landau band. Then $\rho^t_{xx}$ and $\rho^t_{xy}$ defined by

$$\rho^t_{xx(xy)} = \sigma^t_{xx(xy)}/(\sigma^{t\,2}_{xx} + \sigma^{t\,2}_{xy}) \qquad (3)$$

in the 2-1 P-P transition should play the same roles as $\rho_{xx}$ and $\rho_{xy}$ in the 1-0

transition, respectively. [2,3] Figure 2 (a) shows the curve of $\rho^t_{xy}$ at $T$=0.31 K and the curves of $\rho^t_{xx}$ at $T$=0.94-0.31 K in the 2-1 P-P transition, and we can see that $\rho^t_{xx}$ is $T$-independent at $B$=5.5 T, which is the critical point. The temperature-dependeces of $\rho^t_{xx}$ are different on the both sides of the critical magnetic field and hence in this study the 2-1 transition is mapped to the 1-0 transition by the Landau-level addition transformation. In Fig. 2, the region where $\rho^t_{xx}$ remains about the quanitzed value $h/e^2$ covers both sides of the critical point $B$=5.5 T and hence the observed 2-1 transition also satisfies the semicircle law. At the critical point $B$=5.5 T, $\rho^t_{xx}$ is smaller than the expected value $h/e^2$ and hence we can see the failure of the universality of conductivities in the 2-1 transition after transforming $\rho^t_{xx}$ and $\rho^t_{xy}$ to $\sigma_{xx}$ and $\sigma_{xy}$. [2,3,14,16]

While the semicircle law is observed in both the 2-1 and 1-0 transitions, in our study, the universality of critical conductivities is invalid. The broken of the universality of critical conductivities, in fact, does not indicate that the observed transitions are not equivalent. Since such a property is broken in both the 2-1 and 1-0 transitions rather than only in one of them, there should be a common factor to break this property in both transitions. In Fig. 1, in $\rho_{xx}$ the oscillation of $\nu$=3 is unresolved and hence the spin-splitting is small in our study. In the $\nu$=1 quantum Hall state, at $T$=0.94 K the minimum of $\rho_{xx}$ is of the finite value ~0.4 kΩ, which also indicates there is a small spin-splitting. It is suggested by Dolan [16] that the small spin-splitting can result in the deviations on the critical conductivities without breaking the semicircle law, and our observations are consistent with such a suggestion. On the other hand, universal scaling fails for the 2-1 transition but it holds for the 1-0 transition. However, it should be noted that the analysis methods used in these two transitions are different. If we use the scaling fitting method, we could obtain $\kappa$ for the 2-1 transition from the collapse of $\rho_{xx}^t$ with respect to the scaling parameter. With this method, as shown in Fig. 2 (b), in the 2-1 transition we obtained $\kappa$=0.4, which is close to the expected universal value. Therefore, by using the same method of analysis, both 2-1 P-P and 1-0 I-QH transitions are governed by the universal scaling. Our study indicates that evidences to support the universality on different magnetic-field-induced transitions can be found by suitable analysis even when some expected properties are invalid.

The failure of using Eq. (1) to obtain correct $\kappa$ is due to the problem in getting the correct slope of P-P transitions near the critical point. The transition width of P-P transitions could not be easily obtained. To use the inverse of $\max|d\rho_{xy}/dB|$ as the width of the P-P transition in $\rho_{xy}$, it is necessary that the curves of $\rho_{xy}$ can be approximated by the straight line with $\max|d\rho_{xy}/dB|$ as the slope. However, as could be seen in Fig. 3, when the critical point is not near the center of the curve, the slope obtained from $\max|d\rho_{xy}/dB|$ is quite different from the slope in the critical point, and the straight line that used to determine $\max|d\rho_{xy}/dB|$ doesn't necessarily contain the critical point. In fact, we can see in Fig. 3 that the scaling region where the curves $\rho_{xx}^t$ collapse with respect to the scaling parameter $(\nu-\nu_c)/T^\kappa$ does not extend to the whole 2-1 transition, and the scaling region is different from the region where $\rho_{xy}$ can be approximated by a straight line. We found that in $\rho_{xx}$ the scaling region also did not extend to the whole transition region. Our study indicates that in a P-P transition, Eq. (1) is not always suitable to obtain $\kappa$.

In conclusion, well-developed plateau-plateau and insulator-quantum Hall conductor transitions are observed at high fields in the two-dimensional electron system in an AlGaAs/GaAs heterostructure. Both transitions follow the semicircle law, but the universality of critical conductivities is broken and by the conventional scaling analysis the universal scaling is found only in the insulator-quantum Hall conductor transition. The broken of the universality conductivities, which could be due to the small spin-splitting, is observed in both transitions and hence does not indicate the failure of the equivalence between these two transitions. We pointed out that in order to get a correct critical exponent, it is essential that the scaling analysis must be performed near a critical point. And with proper analysis, we found that the P-P transition and the insulator quantum Hall conductor are of the same universal class.

This work is supported by the National Science Council and Ministry of Education of the Republic of China.

**Figure caption**

Fig. 1 The curves of the longitudinal and Hall resistivities $\rho_{xx}$ and $\rho_{xy}$ at the temperature $T$=0.31-0.94K. The inset shows the collapse of $\rho_{xx}/\rho_{xx,B=Bc}$ with respect to $|\nu-\nu_c|/T^{0.38}$ in the 1-0 transition. Here $B_c$, $\nu_c$, and $\rho_{xx,B=Bc}$ are the magnetic field, filling factor, and the value of $\rho_{xx}$ at the critical point in the 1-0 transition, respectively.

Fig. 2 (a) The curve of $\rho^t_{xy}$ at $T$=0.31K and the curves of $\rho^t_{xx}$ at $T$=0.31-0.94K in the 2-1 transition. (b) The collapse of $\rho^t_{xx}$ with respect to $|\nu-\nu_c|/T^{0.4}$ after normalization.

Fig. 3 The curves of $\rho_{xy}$ at $T$=0.31K and 0.94 K in the 2-1 transition. The dot lines are the straight lines with the slope max$|d\rho_{xy}/dB|$. The scaling holds only in the marked region (denoted as the scaling region).

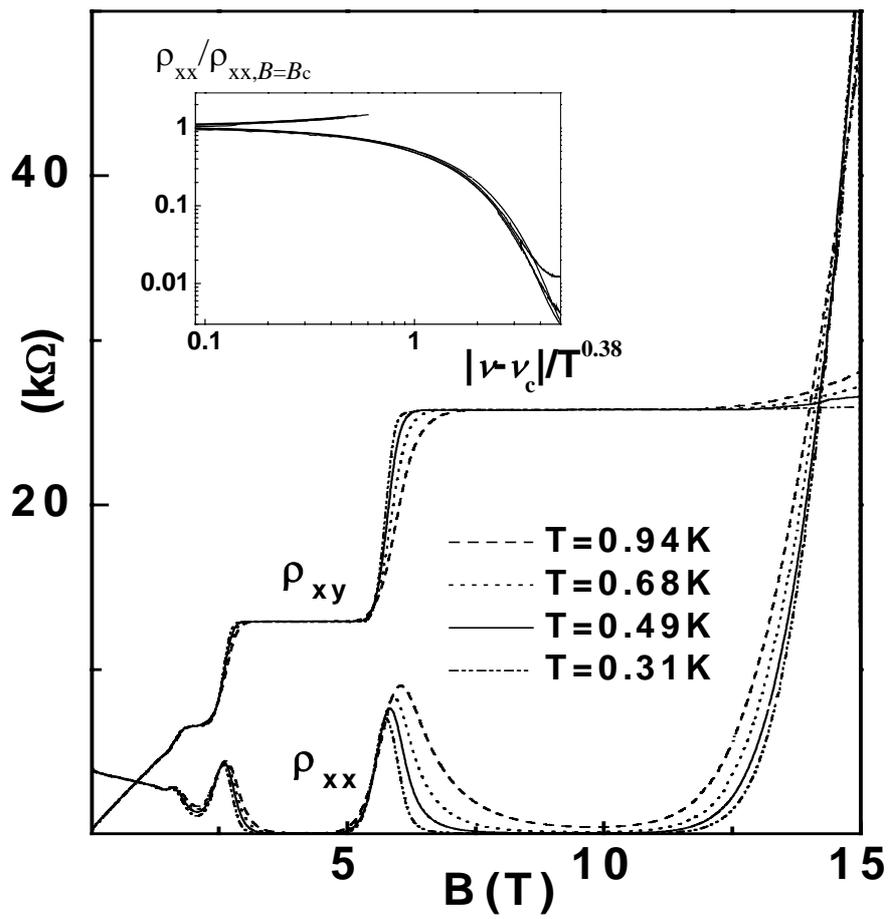

**Fig. 1  C. F. Huang et al.**

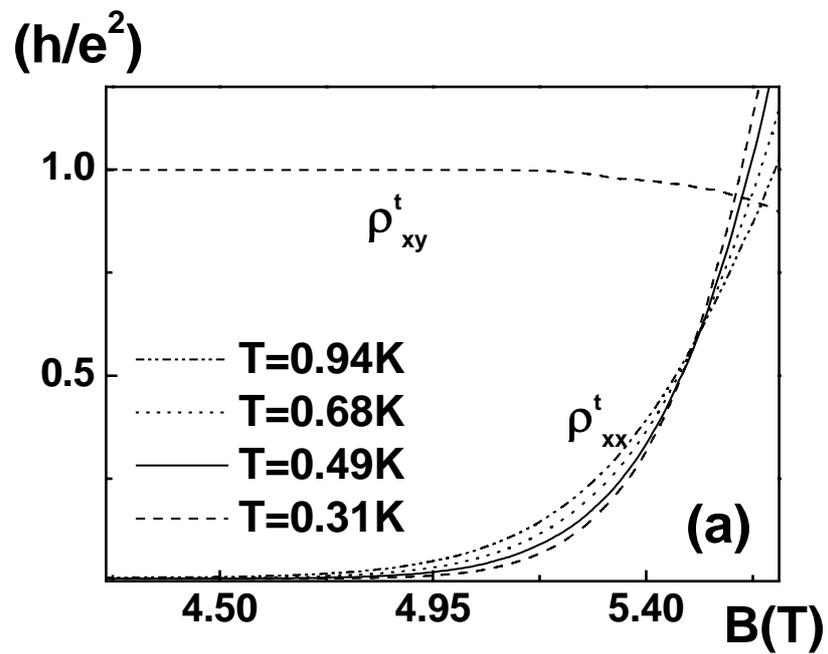

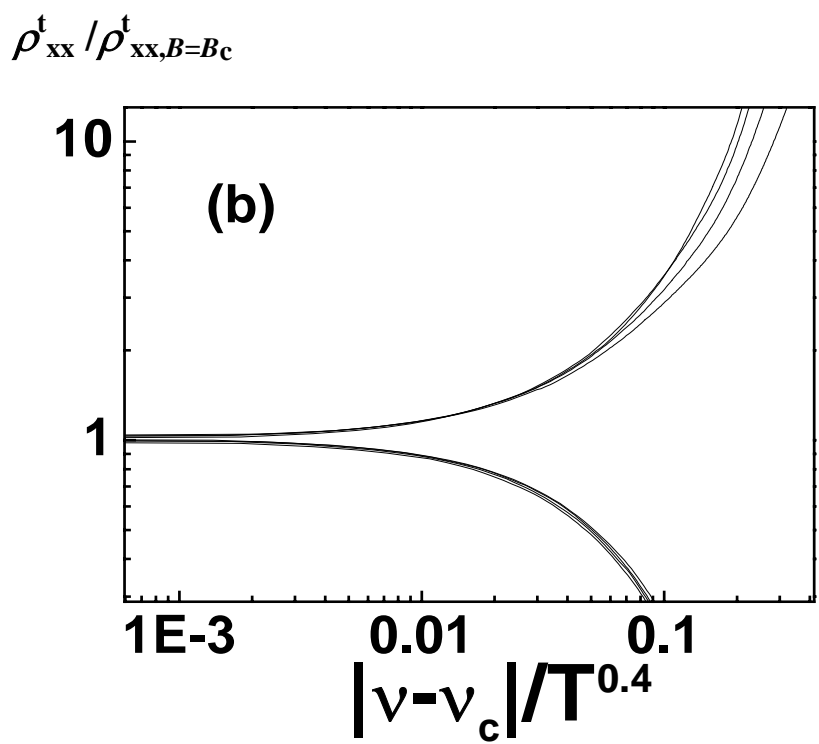

Fig. 2   C. F. Huang et al.

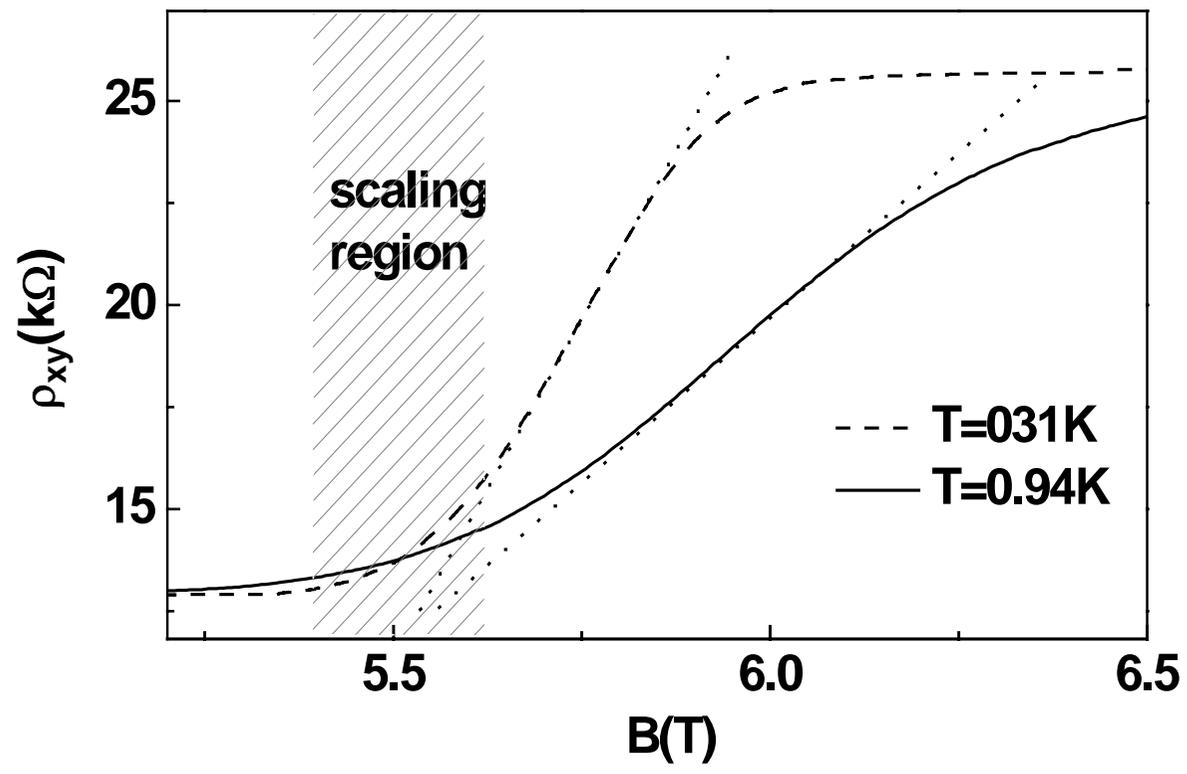

Fig. 3  C. F. Huang et al.

11